\documentclass[aps,showpacs, superscriptaddress, prb]{revtex4-1}
\usepackage{graphicx}
\usepackage{bm}
\usepackage{color}

\begin{document}

\title{Enhancement of electron hot spot relaxation in photoexcited plasmonic structures by thermal diffusion}

\author{F.~Spitzer}
\affiliation{Experimentelle Physik 2, Technische Universit\"at Dortmund, D-44221 Dortmund, Germany}
\author{B.A.~Glavin}
\affiliation{Lashkaryov Institute of Semiconductor Physics, 03028 Kyiv, Ukraine }
\author{V.I.~Belotelov}
\affiliation{Lomonosov Moscow State University, 119991 Moscow, Russia}
\affiliation{Russian Quantum Center, 143025 Skolkovo, Moscow Region, Russia}
\author{J.~Vondran}
\affiliation{Experimentelle Physik 2, Technische Universit\"at Dortmund, D-44221 Dortmund, Germany}
\author{I.A.~Akimov}
\affiliation{Experimentelle Physik 2, Technische Universit\"at Dortmund, D-44221 Dortmund, Germany}
\affiliation{Ioffe Institute, Russian Academy of Sciences, 194021 St. Petersburg, Russia}
\author{S.~Kasture}
\author{V. G.~Achanta}
\affiliation{Tata Institute of Fundamental Research, 400005, Mumbai, India}
\author{D.R.~Yakovlev}
\author{M.~Bayer}
\affiliation{Experimentelle Physik 2, Technische Universit\"at Dortmund, D-44221 Dortmund, Germany}
\affiliation{Ioffe Institute, Russian Academy of Sciences, 194021 St. Petersburg, Russia}

\begin{abstract}
We demonstrate that in confined plasmonic metal structures subject
to ultra-fast laser excitation electron thermal diffusion can
provide relaxation faster than the energy transfer to the lattice.
This relaxation occurs due to excitation of nanometer-sized hot
spots in the confined structure and the sensitivity of its optical
parameters to the perturbation in these regions. Both factors
become essential when the plasmonic resonance condition is met for
both excitation and detection. A pump-probe experiment on
plasmonic gold lattices shows sub-picosecond relaxation with
the characteristic times well-described by a two-temperature
model. The results suggest that dynamical optical response in plasmonic
structures can be tuned by selection of the structural geometry as
well as the choice of wavelength and polarization of the excitation
and detection light.
\end{abstract}


\maketitle

Currently fundamental plasmonic effects are intensely studied in
pure and hybrid metallic structures, which possess also huge
potential for numerous applications ranging from development of
efficient nanolasers with plasmonic cavities to photodynamic cancer
treatment with plasmonic
nanoparticles~\cite{Stockman-SPASER,FocusPlasmonic15}. In this
regard, elucidating and understanding the relaxation dynamics and
heat dissipation in plasmonic elements after excitation by a
femtosecond light pulse is essential~\cite{Vallee-chapter}. In a
homogeneous metal, hot electrons loose their energy by emission of
phonons, following the ultrafast initial stage of photoexcited
carrier thermalization by electron-electron scattering. At room
temperature, in most metals phonon relaxation occurs on time scales
of the order of a picosecond or less~\cite{phonon-time}. In
nanopatterned plasmonic structures, however, the light
absorption becomes strongly inhomogeneous in space and shows in
particular hot spots. Similarly, the reflection and
transmission coefficients depend strongly on the spatial distribution of temperature. As a result, the optical response
of the photoexcited structure is determined not only by electron
cooling through phonon emission, but also by redistribution of
energy in space via electron thermal diffusion (ETD).

In this letter we demonstrate that in plasmonic structures ETD
evolves on timescales comparable to or even shorter than phonon
emission, modifying significantly the sub-ps cooling dynamics as
manifested by the ultrafast optical response. Nanoplasmonic metal
structures are typically embedded in an environment with low thermal
conductivity, so that during times shortly after the optical
excitation they behave as thermally isolated systems with a
correspondingly decelerated heat removal. Nevertheless, we show that in
plasmonic structures which support optical excitation of
nanometer-sized hot spots, ETD becomes feasible on ultrafast sub-ps
time scales and provides a decisive contribution to the electron
relaxation. Our results suggest that dedicated selection of material
and geometry of a confined plasmonic structure allows tuning of
its dynamical optical response by controlling the ETD process parameters.

\begin{figure}[h]
    \centering
    \includegraphics[width=0.5\linewidth]{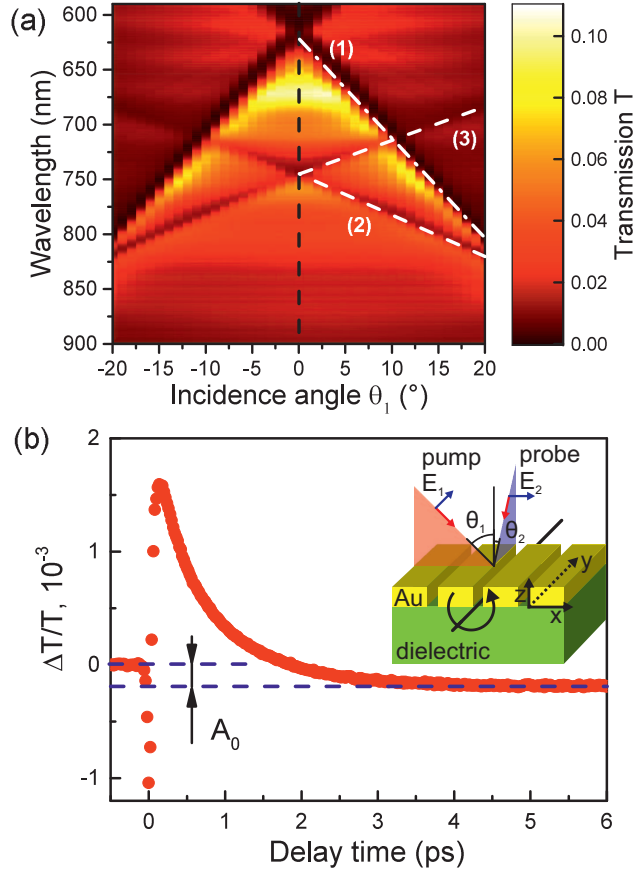}
    \caption{(a)
Contour plot of transmission $T$ as function of wavelength
and incidence angle $\theta_1$ in PC with stripe width $d_x=460$~nm.
The spectra were taken with a white light source in $p$-pol.
Dashed lines indicate SPP dispersion at Au-air (1) and Au-dielectric (2,3) interfaces. (b) Differential
transmission transient $\Delta T/T$ measured at incidence angle of
$\theta_1=17^\circ$ with $p$-polarized pump and $s$-polarized probe.
Inset shows experimental geometry.}
    \label{fig1}
\end{figure}

For the ETD studies we used a model system of a plasmonic
nanostructure: lateral gold gratings on top of a dielectric garnet
layer, as schematically shown in Fig.~\ref{fig1}. These structures
can be considered as plasmonic crystals (PCs) which support surface
plasmon polaritons (SPPs). The grating period is comparable to the
wavelength of light in the visible/near-infrared spectral range which
allows coupling of the far field radiation to the SPPs and vice
versa. The transmission spectra in Fig.~\ref{fig1}(a) show clear
resonances which correspond to propagating SPPs at the metal-air and
metal-dielectric interfaces. The investigated PCs combine two prime
features in a single structure: high tunability of the plasmonic
resonances and heat confinement in each of the grating stripes. The
tunability is achieved by variation of the incidence angle and
proper choice of the polarization of the excitation light pulse.
This enables one to generate different hot spot distributions within
a stripe and to modify the ETD flow. The thermal isolation is ensured by deposition of the
PC on the dielectric substrate with low thermal conductivity. Most of the results presented
below were obtained for a structure with stripe width $d_x =
460$~nm, stripe height $d_z = 130$~nm, and grating period
$a=590$~nm. In addition, measurements were performed on structures
with $d_x=290$~nm, $d_z=80$~nm, and $a=400$~nm, to study the scaling
of the ETD relaxation with the lateral PC size. Additional details
on the PC parameters and sample preparation are given in
Refs.~\cite{Belotelov-09, Brueggemann-12}.

For measuring the dynamics of the PC optical response to pulsed
excitation we implemented a pump-probe technique based on a Ti-Sa
oscillator with a repetition frequency of 80~MHz emitting a
broad-band spectrum centered at 800~nm wavelength with a full width
at half maximum of about 80~nm. The optical pulses were
chirp-compensated before entering the sample, providing an overall
time resolution below 40~fs. The pump beam hit the sample in the
plane perpendicular to the slits, while the probe beam was incident
in the slit plane. The incidence angles of the pump $\theta_1$, and
the probe $\theta_2$, could be tuned by rotating the sample around
the axis parallel to the slits. All measurements were done at
room temperature (300~K). An example of a differential transmission
transient $\Delta T(t)/T$ for $\theta_1 = \theta_2 =17^\circ$ is
shown in Fig.~\ref{fig1}(b). The negative spike at zero delay is due
to instantaneous non-linear response of the
dielectric~\cite{Pohl-12}. It is followed by a rapid rise of the
transmission with a time constant of about 50~fs which is attributed
to the dynamical response of the imaginary part of the dielectric
function due to the hot electron thermalization by electron-electron
scattering right after the excitation pulse. The subsequent decay of
the differential transmission during about a ps is due to
electron-phonon relaxation, affecting mainly the real part of the
dielectric function. Finally, there is an offset $A_0$ which
originates from an increase of the lattice temperature by the
optical excitation. In the following we concentrate on the sub-ps
dynamics and substract $A_0$ from the signals.

\begin{figure}[h]
    \centering
    \includegraphics[width=0.5\linewidth]{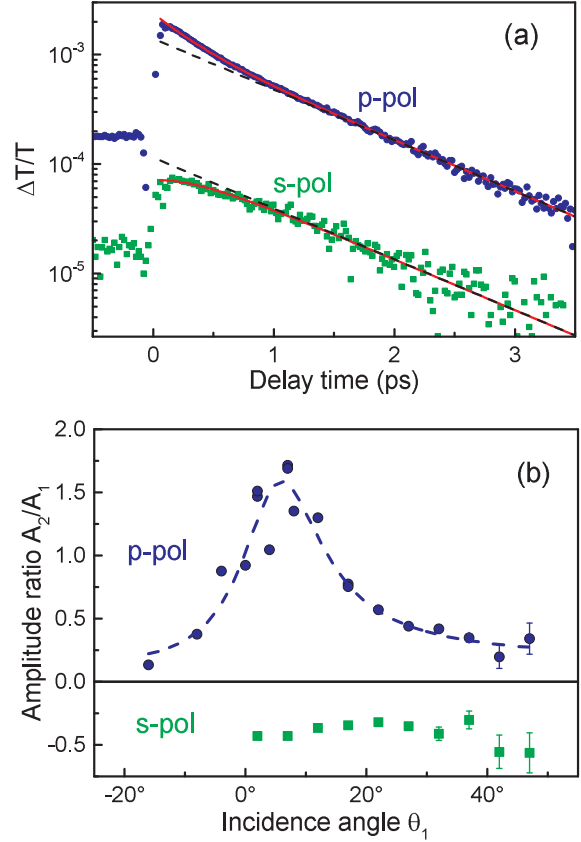}
    \caption{
(a) Differential transmission transients taken at $\theta_1 =
17^\circ$ for $p$- and $s$-polarized pump after substraction of
$A_0$. Solid lines are fits with double exponential decay with
$\tau_1=940$~fs and $\tau_2=300$~fs. (b) Dependence of $A_2/A_1$ on
incidence angle $\theta_1$ for the two polarizations. The data are
presented for a PC with $d_x=460$~nm. Dashed lines in (a) and (b) are guides to the
eye.}
    \label{fig2}
\end{figure}

Fig.~\ref{fig2} summarizes the data for different excitation
conditions with plasmonic probe. Fig.~\ref{fig2}(a) compares two transients for which in
one case the pump beam polarization is plasmonic ($p$-pol, SPPs are
excited) and non-plasmonic in the other case ($s$-pol, no SPP
excitation). By switching from plasmonic to non-plasmonic excitation we observe
a strong decrease of the signal due to less efficient excitation of
hot carriers, in agreement with our previous work~\cite{Pohl-12}.
However, the most striking feature is the observation of double
exponential decays of the optical transmission change $\Delta T$
induced by the pump. We note that the transients scale linearly with
pump excitation power (up to 0.1~mJ/cm$^{-2}$) which suggests that
this behavior  represents a linear response. In
order to quantify the signal contributions we use the function
\begin{equation}
\label{eq:fit} \Delta T(t)/T = A_1\exp
\left(-t/\tau_1\right) + A_2\exp
\left(-t/\tau_2\right),
\end{equation}
to fit the decay with the first (second) term describing the slow
(fast) component. Variation of polarization and incidence angles of
pump and probe results in changes of the amplitudes $A_{1,2}$ only,
while the times $\tau_{1,2}$ remain constant for a given structure.
In the PC with $d_x=460$~nm fitting the experimental data provides
$\tau_1=940$~fs and $\tau_2=300$~fs. Further, for plasmonic excitation $A_2/A_1$
is positive and strongly depends on the
incidence angle. Otherwise, for
$s$-polarized pump the ratio is negative and practically does not
depend on $\theta_1$ (see Fig.~\ref{fig2}(b)).
Similar measurements on the PC with $a=400$~nm show a double-exponential decay
with $\tau_1 = 940$~fs and $\tau_2=220$~fs.

The two-temperature model~\cite{2T} used for analysis of the cooling
dynamics and its manifestation in the pump-probe experiment are
given in the appendix~\ref{A1}. The evolution of the
electron temperature $T_e$ has a complex behavior which can be
analyzed by decomposing the spatial distribution of
$T_e$ into the eigenmodes of the thermal problem. Each eigenmode is
characterized by an individual relaxation time, resulting in general
in a  multi-exponential behavior. The amplitude weight
of each eigenmode is determined by its contribution to the initial
spatial distribution  of $T_e$, initiated by the pump pulse. For
the rectangular PC stripe, these eigenmodes are the harmonic Fourier components
labeled by the indices $n$ and $m$
for the $x$- and $z$-directions (see appendix \ref{A1}).
Their initial amplitudes $B_{nm}$ are given by their overlap integrals with
the initial distribution of $T_e$. For the perturbation of the
optical transmission by the pump $\Delta T$, we obtain
%in that way
\begin{equation}
\label{eq:perm}
\Delta T =
\frac{d\varepsilon}{d T_e} \sum_{n,m=0}^\infty k_{nm} B_{nm} \exp\left(- t/\tau_{nm} \right).
\end{equation}
Here we assume that the perturbation of the dielectric function
$\varepsilon$, which is proportional to the rise of $T_e$, is real,
and $k_{nm}$ are the perturbations of the
transmission in response to the change of $\varepsilon$ having the spatial structure
of the $(n,m)$ harmonic. The decay
constant of each component $\tau_{nm}^{-1} = \tau_{eph}^{-1} +
\frac{n^2}{\tau_x} + \frac{m^2}{\tau_z}$, where $\tau_{eph}$ is the
time of electron energy relaxation due to phonon emission, and
$\tau_{x,z} = C_e d_{x,z}^2 \pi^{-2} \kappa_e^{-1}$. Here $\kappa_e$
and $C_e$ are the thermal conductivity and thermal capacity of the
electrons, respectively. Note that Eq.(\ref{eq:perm}) reflects the
main features of the relaxation process. The spectrum of relaxation
times is determined by geometry and material of the structure and
not directly related with its electromagnetic properties. On the
other hand, optical access to a particular eigenmode depends on
the possibility to couple the corresponding spatial profile of the perturbation to the laser radiation,
both in excitation and detection. This coupling is
determined by the plasmonic properties of the structure. It can be controlled by
tuning wavelength, polarization and incidence angle of the laser.
In our case, the plasmonic resonance enhances the amplitudes of
particular harmonics with $n,m \neq 0$ such that they are comparable
to the $n,m=0$ contribution. If the probe is also in the plasmonic
resonance band, its sensitivity to the spatially-inhomogeneous
pattern may be enhanced for particular harmonics with $n,m \neq 0$.

The double exponential decay in Fig.~2 suggests that mostly two
Fourier components contribute to the thermal perturbation. Let us
provide some estimates in this respect.
We assume $\kappa_e =318$~WK$^{-1}$m$^{-1}$, $C_e=
\gamma T_e$, and $\gamma = 68$~JK$^{-2}$m$^{-3}$
\cite{mat-param1,mat-param2}, from which we obtain $\tau_x=1.37$~ps,
$\tau_z=0.11$~ps. Obviously $\tau_z$ is very short and comparable to
the electron scattering time \cite{Vallee-chapter}. This means that a thermal conduction
approach can hardly be applied to modeling of the electron energy
transfer along the $z$-direction. As we are interested in the
electron relaxation after the initial stage lasting a few tens of fs, we
can analyze the subsequent dynamics using the ETD approach solely for the $x$-direction
considering the $m=0$ harmonics.
Another peculiarity of the PC is that even for finite incidence
angle the pump excitation pattern is close to an even function of
$x$ with respect to the center of the stripe. As a result, the most
important contributions arise from terms with even $n$,
corresponding to even harmonics. From calculations of the amplitudes
$B_{nm}$ we conclude that the essential Fourier components are those with
$n=0$ and $n=2$, both with $m=0$. This is in line with the
double-exponential behaviour in Fig.~2, so that we can identify
$\tau_{1,2}$ with $\tau_{00}$ and $\tau_{02}$. To obtain
quantitative agreement with the measured values, we have to assume a
somewhat reduced $\kappa_e$ of about 250~WK$^{-1}$m$^{-1}$. This
reduction may be caused by additional scattering of electrons at the
surface and interfaces of the stripes \cite{gold-th-transport}. For
the PC with $d_x=290$~nm, theoretical and experimental values
match for $\kappa_e$ about 150~WK$^{-1}$m$^{-1}$. The stronger
reduction of $\kappa_e$ in this case may be due to the smaller
stripe thickness, enhancing the role of surface electron scattering.

\begin{figure}[h]
    \centering
    \includegraphics[width=0.7\linewidth]{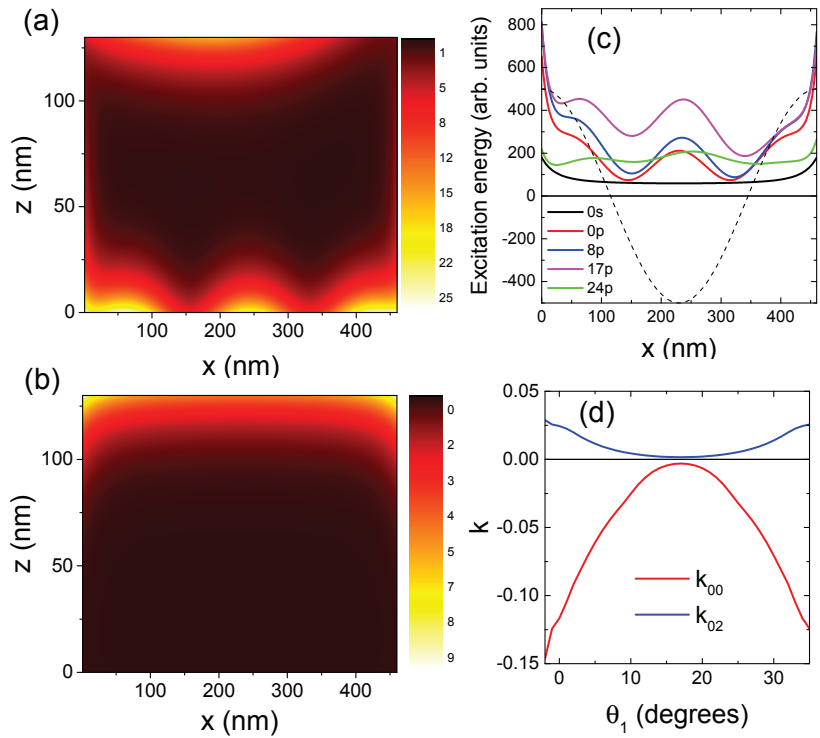}
    \caption{(a)
Contour plots of spatial distribution of the absorbed pump energy in
PC stripe (arbitrary units). (a) and (b) correspond to excitation by
$p$-polarized light ($\theta_1 =17^{\circ}$) and $s$-polarized
light, respectively. Bright areas correspond to hot spots. In (c)
the $x$-dependence of absorbed energy averaged over the
$z$-direction is shown for various incidence angles at
$p$-pol. and normal incidence for $s$-pol. For
comparison also the harmonic with $n$=2 in the Fourier analysis of
the  thermal conduction problem, which provides fast relaxation with
time constant $\tau_{02}$, is shown. In (d) the angle dependence of
coefficients $k_{00}$ and $k_{02}$ is shown.}
    \label{fig3}
\end{figure}

The amplitudes $A_{1,2}$  deserve special
attention. To that end, we show in Figs.~\ref{fig3}(a) and (b) the
spatial patterns of absorbed energy in a single stripe for plasmonic
($p$-pol, top) and non-plasmonic ($s$-pol, bottom)
excitation. The bright areas indicate hot spot formation. The
absorbed energy is proportional to $\left|\bm{E}\right|^2$, where
$\bm{E}$ is the complex amplitude of the electric field vector of
the optical pump wave in the metal from which the initial electron
temperature distribution arises. The spatial distribution of
$\bm{E}$ was calculated applying a rigorous coupled-wave analysis
\cite{RCWA}. The excitation pattern for plasmonic polarization is
shown for the incidence angle $\theta_1 = 17^{\circ}$, which is
close to the plasmonic resonance; for non-plasmonic excitation it is
practically independent on $\theta_1$.

For plasmonic excitation a pronounced hot spot structure is present
scaling spatially with the wavelength of the uniform film SPPs,
excited by the pump in the perforated structure. In
Fig.~\ref{fig3}(c) the absorbed energy (averaged over the
$z$-direction) is shown for different incidence angles $\theta_1$
demonstrating the absorption resonance for $\theta_1 = 17^{\circ}$.
The hot spots contribute considerably to the modes with $n \neq 0$.
Obviously, the excitation spatial profile can not be described by the combination
of the uniform, $n=0$, and $n=2$ harmonics (the latter is shown  in Fig.~\ref{fig3}(c) by the
dashed line). However, for the harmonics with $n>2$, $\tau_{nm}$ is very short. Similar to the
case of vertical relaxation, the corresponding harmonics probably relax in a non-diffusive way
being unresolvable for pump-probe delays exceeding a few tens of fs.
For non-plasmonic excitation (shown for normal incidence
only) the absorption is much weaker. However, also here hot spots
are formed at the sharp edges of the rectangular stripes, which can
be well described by the $n=0$ and $n=2$ terms in the Fourier
series.

Next, we have to consider how the local variation of the dielectric
fucntion translates into changes of transmitted probe intensity. The
calculated angle dependencies of the coefficients $k_{00}$ and
$k_{02}$ are shown in Fig.\ref{fig3}(d). The results explain the
negative sign of the ratio $A_2/A_1$ observed for non-plasmonic
excitation. Indeed, the coefficients $k_{00}$ and $k_{02}$ have
opposite signs for all considered angles. $B_{00}$, being the
average absorbed excitation energy, is per se positive. Localization
of the hot spots near the stripe edges for non-plasmonic pump ensures
$B_{02}>0$. For the plasmonic
configuration, we do not achieve quantitative agreement of
experiment and theory. The experimental data suggest negative
$B_{02}$, which would occur for intense hot spots in the middle of
the stripes. They do appear in the calculations due to the
excitation of the plasmonic resonance (see Fig.~3(a)), but the
calculated hot spots near the wall edges provide larger
contributions for incidence angles of $\theta_1=$ 0, 8, and
17$^{\circ}$. Thus $B_{02}$ should be positive there as well.
According to our calculations, the  absorbed energy near the
edges is strongly sensitive to the stripe shape. Therefore, we
tentatively attribute the discrepancy of experiment and theory to
deviation of the stripe shape from the assumed rectangular one.

The developed model allows us to estimate the role of ETD also in
plasmonic structures with a shape differing from a stripe and/or
fabricated from a metal other than gold.  ETD becomes important if its characteristic
relaxation rate $\pi^2 \kappa_e C_e^{-1} d^{-2}$, where $d$ is the
typical size of the hot-spot pattern in the plasmonic structure, is
comparable to or exceeds $\tau_{eph}^{-1}$. From setting the two
rates equal, we can determine the required geometrical dimension of
the plasmonic structure that needs to be undercut in order to reach
these conditions, $d_{th}=\pi \sqrt{\kappa_e C_e^{-1}\tau_{eph}}$,
which establishes a threshold for ETD becoming efficient. In table
\ref{tab:matdata} we list the  $d_{th}$ as well as the $\tau_{eph}$
for various metals at 300~K. By tailoring material and geometry of
the structure it is possible to obtain ETD  relaxation faster than that due to phonon emission.
In this case also shape design is
important to provide a strongly inhomogeneous optical excitation
pattern.

We would like to address the recent paper by Harutyunyan et al. \cite{Govorov-2015} in which optical excitation of hot spots in
plasmonic nano-discs was attributed to the appearance of a short
optical response in differential reflection transients. However, the
relaxation time constant could not be measured due to insufficient
time resolution. Moreover, the role of ETD was not discussed as
possible origin of the fast relaxation component.

\begin{table}
\begin{tabular}{|ccccccccc|}
\hline
                                                             & Au   & Ag    & Cu  & Al    & Ni   & Pt    & W    & Fe   \\ \hline
%$\kappa_e \left(\frac{W}{m K}\right) $     &317   &429   &401  &237  &90.7 &71.6  &174  &80.2 \\
$\tau_{eph}$ (ps)                                  &1       &0.5    &0.3   &0.2 &1     &0.2    &0.06 &0.7 \\
$d_{th}$ (nm)                                       &400   &350   &200   &100 &50    &25    &60    &50 \\ \hline
\end{tabular}
\caption{Electron-phonon cooling time and $d_{th}$ for different
metals. The material parameters are taken from \protect
\cite{phonon-time,mat-param1,mat-param2,iron}} \label{tab:matdata}
\end{table}

In conclusion, we demonstrate that the relaxation in a photoexcited
plasmonic grating can be enhanced by thermal diffusion of hot
electrons on a sub-picosecond time scale, essentially below the
value established for electron cooling by phonon emission. This
fast relaxation is possible due to formation of nanometer-sized hot
spots under ultra-fast optical excitation if the plasmonic resonance
condition is met. Our analysis based on the two-temperature model
provides good agreement with the experimentally observed relaxation
times and allows us to formulate a geometric criterion for the
importance of thermal diffusion which can be adjusted through the
plasmonic structure material and design.

\appendix

\section{Theoretical consideration of thermal relaxation and transient optical response\\ in metallic grating}\label{A1}
Here, we theoretically consider the thermal relaxation
of electrons in a metallic grating and the consequences on the
transient optical response of the grating. The relaxation of
photoexcited electrons in metals occurs in several stages. Commonly,
the fastest stage implies thermalization of electrons resulting in a
quasi-equilibrium electron distribution with the electron
temperature exceeding that of the crystal lattice. This relaxation
is due to efficient electron-electron scattering, and the
corresponding times in metals are in the range of a few tens of fs
(for gold, this value is claimed to be about 30 fs). Once
thermalization is reached, the most universal process is cooling of
electrons due to emission of phonons, i.e. transfer of the excess
energy to the crystal lattice. For gold, the characteristic time of
this process, $\tau_{eph}$ , is about 1 ps. For systems with a
spatially inhomogeneous distribution of photoexcitation these
processes are complemented by spatial electron energy transfer
towards a balanced distribution. It has a quasi-ballistic character
until electron thermalization is reached, switching to thermal
diffusion afterwards. After the thermalization, the electron-lattice
energy transfer can be described by the two-temperature model~\cite{2T},
which describes the electron-lattice energy balance where thermal
diffusion occurs mainly due to electrons possessing superior thermal
conductivity in comparison to the lattice. The underlying equations
for electron and lattice tempearture variation with time are:
\begin{eqnarray}
\label{eq:2T}
C_e \frac{\partial T_e}{\partial t}&=&\nabla \cdot \left( \kappa_e \nabla T_e\right) -G(T_e,T_l) +F (\bm r, t), \nonumber \\
C_l \frac{\partial T_l}{\partial t}&=&G (T_e,T_l).
\end{eqnarray}
Here $C_{e,l}$ are the electron and lattice heat capacity,
$\kappa_e$ is the electron heat conductivity, $G$ is the
electron-phonon coupling factor, and $F$ describes the external
power input into the electronic subsystem. The lattice thermal
conductivity is neglected. The common approximation valid for
$T_{e,l}$ exceeding the Debye temperature is $G=g (T_e - T_l)$ with
the temperature-independent $g$. In the linear-response regime,
corresponding to small excitation fluence, we can neglect the
temperature dependence of $\kappa_e$, $C_{e,l}$. It is important
that in infinite or semi-infinite structures the thermal diffusion
is not described by a characteristic time corresponding to an
exponential decay, but rather obeys a power-like law. The situation
is essentially different for spatially confined structures. Taking
into account that the excitation of plasmonic structure takes place
very fast, $F$ can be assumed to be a $\delta$-like function of
time. The solution of Eq.(\ref{eq:2T}) can be written in terms of
the initial electron temperature distribution $\Delta T_0 (\bm{r})$.
After simple algebra we obtain that this solution can be presented
as a series
\begin{eqnarray}
\label{eq:2T-sol}
T_e&=&\sum_{n=0}^\infty \left( B_n^{(1)}\exp \left(-t/\tau_n^{(1)}\right) + B_n^{(2)} \exp \left(-t/\tau_n^{(2)}\right)\right) T_n (\bm{r}) \\
T_l&=&\sum_{n=0}^\infty \left( \frac{B_n^{(1)}}{1- \tau_{eph}/(\tau_n^{(1)} \xi)} \exp \left(-t/\tau_n^{(1)}\right) + \frac{B_n^{(2)}}{1- \tau_{eph}/(\tau_n^{(2)}/\xi)} \exp \left(-t/\tau_n^{(2)}\right)\right) T_n (\bm{r}). \nonumber
\end{eqnarray}
Here $\tau_{eph} = C_e/g$ is the electron-phonon energy relaxation
time, $\xi = C_e/C_l$, and $T_n (\bm r)$ are the eigenmodes of the
Helmholtz equation
\begin{equation}
\label{eq:helm}
\nabla^2 T_n + q_n^2 T_n = 0
\end{equation}
with the boundary condition of a zero gradient component of $T_n$
normal to the surface, corresponding to the boundary condition of a
thermally isolated structure. This problem provides eigenvalues
$q_n^2$, each of which enter into the following two decay rates
\begin{equation}
\label{eq:rate} \frac{1}{\tau_n^{(1,2)}} = \frac{1}{2} \left(
\frac{1+\xi}{\tau_{eph}} + \frac{1}{\tau_n} \pm
\sqrt{\left(\frac{1}{\tau_n} + \frac{1}{\tau_{eph}}\right)^2 -2\xi
\frac{1}{\tau_{eph} \tau_n} +\xi^2 \frac{1}{\tau_{eph}^2}} \right),
\end{equation}
where $\tau_n = \kappa_e (C_e q_n^2)^{-1}$ is the characteristic
thermal diffusion relaxation time of the $n$-th mode. The
coefficients $B_n^{(1,2)}$ are determined by the following
expression:
\begin{equation}
\label{eq:ini}
B_n^{(1,2)} = \frac{B_n \tau_n^{(2,1)} (\xi \tau_n^{(1,2)}-\tau_{eph})}{(\tau_n^{(1,2)} - \tau_n^{(2,1)})\tau_{eph}},
\end{equation}
where the $B_n$ are the coefficients in the expansion $\Delta T_0
({\bm r}) = \sum_n B_n T_n({\bm r})$.

For metals usually $\xi \ll 1$. In this limit we obtain
\begin{eqnarray}
\label{eq:rate-aprox}
\frac{1}{\tau_n^{(1)}} &=& \frac{1}{\tau_{eph}} +\frac{1}{\tau_{n}} \\
\frac{1}{\tau_n^{(2)}} &=& \xi \frac{1}{\tau_{eph}}
\left(1- \frac{1}{2}\frac{\tau_n}{\tau_{eph}+ \tau_n}\right) \nonumber \\
B_n^{(1)} &=& B_n, B_n^{(2)} = - B_n \frac{\tau_n^{(1)}}{ \tau_n^{(2)}} \nonumber
\end{eqnarray}
These expressions reflect the fact that due to the big difference of
electron and lattice heat capacities hot electrons, at the first
stage, lose their energy due to phonon emission and thermal
diffusion reaching a quasi-equilibrium state with the lattice. This
stage is characterized by the times $\tau_n^{(1)}$. Then, much
slower relaxation occurs on the time scale $\tau_n^{(2)}$, where
both electron and lattice temperatures finally reach a uniform
state. Since we are interested in the analysis of the fast stage of
relaxation, we disregard the contributions of the $B_n^{(2)}$-terms
to the electron temperature evolution.

Let us provide explicit results for the thermal relaxation within a
stripe of a plasmonic grating. We chose the coordinate system such
that the stripe is within the region $0<z<d_z$, $0<x<d_x$. From the
general expressions for $\Delta T_0 (x,z)$ we obtain
\begin{equation}
\label{eq:en-sol}
\Delta T_e \equiv T_e-T_l =\sum_{n,m=0}^\infty B_{nm} \exp\left(- t/\tau_{nm}\right)
\cos\frac{\pi n x}{d_x} \cos\frac{\pi m z}{d_z},
\end{equation}
where
\begin{eqnarray}
\label{eq:lambda}
\frac{1}{\tau_{nm}} &=& \frac{1}{\tau_{eph}} + \frac{n^2}{\tau_x}+ \frac{m^2}{\tau_z}, \nonumber \\
\tau_{x,z} &=& \frac{C_e d_{x,z}^2}{\pi^2 \kappa}.
\end{eqnarray}
The coefficients $B_{nm}$ are determined by $\Delta T_0 (x,z)$:
\begin{equation}
\label{eq:en-sol}
B_{nm} =\frac{\xi_n \xi_m}{d_x d_z} \int_0^{d_x} dx \int_0^{d_z} dz
\Delta T_0 (x,z) \cos\frac{\pi n x}{d_x} \cos\frac{\pi m z}{d_z},
\end{equation}
with $\xi_i =2$ for $i=0$ and $\xi_i =1$  otherwise. As we see the
relaxation of nonequilibrium electrons has a multi-exponential
character, with the exponents strongly dependent on the geometry of
the structure (in the particular case of a stripe, on the dimensions
of its cross-section). For actual structures $\tau_{x,z}$ can be as
small as a few hundreds of fs.

The mechanism of the optical response to electron heating is based
on perturbation of the metal dielectric permittivity by heating of
the electrons~\cite{Vallee-chapter}. In the linear response regime we can assume that
this perturbation is proportional to $\Delta T_e$. As a result, the
perturbation of the dielectric perimittivity is given by
\begin{equation}
\label{eq:perm}
\Delta \varepsilon =\frac{d\varepsilon}{dT_e}
\sum_{n,m=0}^\infty B_{nm} \exp\left(- t/\tau_{nm} \right)
\cos\frac{\pi n x}{d_x} \cos\frac{\pi m z}{d_z},
\end{equation}
where the derivative is calculated for ambient temperature. In
general, it is a complex number, and we present it in the form
$d\varepsilon /dT_e=\zeta \exp(i \varphi )$. To calculate the
perturbation of the optical transmission of the system, $\Delta
T_{opt}$, we introduce the coefficients $k_{nm}$, defined as
\begin{equation}
\label{eq:knm}
k_{nm} = \frac{d\Delta T_{nm}}{d\varepsilon_{nm}},
\end{equation}
where $\Delta T_{nm}$ is the perturbation of transmission induced by
that of the dielectric permittivity $\varepsilon_{mn} \exp(i \varphi
) \cos\frac{\pi n x}{d_x} \cos\frac{\pi m z}{d_z}$. In the linear
response regime we have
\begin{equation}
\label{eq:perm}
\Delta T =\zeta \sum_{n,m=0}^\infty k_{nm}B_{nm} \exp\left(-t/\tau_{nm} \right),
\end{equation}

\begin{acknowledgments}
The authors are grateful to Natalia Del Fatti and Fabrice Vall\'ee for valuable discussions.
The work was supported by the Deutsche Forschungsgemeinschaft (Projects ICRC TRR
160 and AK40/7-1), Volkswagen Foundation (Grant 90418), the Russian Foundation for Basic Research (Grant No. 16-02-01065), Russian President Grant (ID-5763.2015.2).
VGA acknowledges Department of Science and Technology (DST), India.
\end{acknowledgments}

\end{document}